\documentclass{article}

\usepackage{microtype}
\usepackage{graphicx}
\usepackage{booktabs}
\usepackage{array}
\usepackage{tabularx}
\usepackage{makecell}
\usepackage{hyperref}
\usepackage{xurl}

\hypersetup{
  pdftitle={Macro-Prudential AI Governance: A Two-Layer Early Warning and Response System for Frontier AI},
  pdfauthor={Pranav Mehta}
}

\usepackage[accepted]{icml2026}

\usepackage{amsmath}
\usepackage{amssymb}
\usepackage{mathtools}
\usepackage{amsthm}

\usepackage[capitalize,noabbrev]{cleveref}

\theoremstyle{plain}

\theoremstyle{definition}

\theoremstyle{remark}

\icmltitlerunning{Macro-Prudential AI Governance}

\usepackage{tikz}
\usetikzlibrary{arrows.meta, positioning, calc, backgrounds, fit}
\definecolor{laBlue}{HTML}{345088}
\definecolor{laBlueTint}{HTML}{EEF2F8}
\definecolor{laBlueFaint}{HTML}{9BB0CF}
\definecolor{lbGold}{HTML}{97632E}
\definecolor{lbGoldTint}{HTML}{F6F0E6}
\definecolor{busDark}{HTML}{20242B}
\definecolor{descGray}{HTML}{5B6068}

\newcommand{\mewrscell}[2]{%
  \begin{minipage}{3.1cm}\raggedright
    \textbf{\small #1}\\[2pt]
    {\scriptsize\itshape\color{descGray} #2}
  \end{minipage}}

\begin{document}

\twocolumn[
\icmltitle{Macro-Prudential AI Governance: A Two-Layer Early Warning\\
and Response System for Frontier AI}

\begin{icmlauthorlist}
\icmlauthor{Pranav Mehta}{ind}
\end{icmlauthorlist}

\icmlaffiliation{ind}{MATS Research, Berkeley, CA, USA}

\icmlcorrespondingauthor{Pranav Mehta}{pranav.mehta@columbia.edu}

\icmlkeywords{technical AI governance, macro-prudential regulation, early warning, frontier AI, Basel III, systemic risk}

\vskip 0.3in
]

\printAffiliationsAndNotice{}
\begin{abstract}
Frontier-AI governance today faces a problem structurally analogous to the one banking regulation faced \emph{pre}-2008, and which \emph{post}-2008 reforms (Basel~III, Dodd--Frank) have since addressed. Two gaps recur: discovering a risk is not tantamount to acting on it, and individual-model review is unlike managing correlated build-up across the sector. Drawing on the Basel~III framework and the U.S.\ financial-stability architecture, I propose a macro-prudential early warning and response system ("MEWRS") for internal frontier AI. These are systems deployed for labs' own internal research, testing, and production workflows, as distinct from externally released products. Layer~A adapts the finder-coordinator-defender early-warning model to route structured reports on dual-use capabilities, autonomy indicators, and security compromises through a government clearinghouse to domain-specific defender working groups. Layer~B calibrates operational controls via three quantitative buffer metrics, namely Effective Compute-at-Risk (ECAR), Cumulative Red-Team Hours (CRTH), and an Alignment Robustness Score (ARS), so that faster capability scaling automatically triggers stronger safeguards, analogously to how risk-weighted assets drive capital ratios under Basel~III. I outline the reporting schema, map six Basel~III mechanisms onto AI-governance analogues, identify seven failure modes with concrete mitigations, and sketch an exercise-based validation plan. MEWRS is designed to detect correlated risk build-ups across the frontier-AI sector and create pre-committed off-ramps before a cascade unfolds. 
\end{abstract}
\section{Introduction}
\label{sec:intro}

Post-2008 financial regulation solved an adjacent pair of problems with two families of mechanisms. \emph{Micro-prudential} tools (capital buffers, stress testing, disclosure) strengthen individual institutions. \emph{Macro-prudential} tools (counter-cyclical buffers, systemic-institution tiering, resolution ``living wills'') target correlated exposures and self-reinforcing dynamics across the sector \citep{bcbs2017basel, fsb2011sifi}. Current frontier-AI governance work, including dangerous-capability evaluations \citep{shevlane2023model}, responsible scaling policies \citep{anthropic2026rsp, openai2025preparedness}, and structured frontier-regulation proposals \citep{anderljung2023frontier}, maps cleanly onto the micro-prudential side: each targets the safety of individual models or individual release decisions. The macro-prudential toolkit, designed for correlated risk build-up across institutions, is underdeveloped, leaving the sector exposed to sector-level failure modes. For example, a jailbreak technique discovered against one lab's agentic system may generalize to structurally similar systems at other labs before any single lab's internal review cycle can respond (see \S\ref{sec:threats}).

A parallel gap exists \emph{within} the development pipeline. A growing body of research has developed tools for model evaluations, red-teaming, and early-warning indicators \citep{shevlane2023model, anderljung2023frontier}, though their formalisation in regulation remains limited --- the principal exception being the EU AI Act's systemic-risk obligations for general-purpose AI models (model evaluation, adversarial testing, and serious-incident reporting under Art.~55), operationalised through the General-Purpose AI Code of Practice finalised in July 2025 \citep{euaiact2024, ec2025gpaicode}. These tools have become more sophisticated, even as the underlying task of reliable capability elicitation has become harder at frontier scale, with evidence of sandbagging and evaluation-awareness in frontier models \citep{vanderweij2024ai, meinke2024frontier}. Much less has been built to ensure that discovery leads to \emph{action}: that a capability finding or security incident reaches the right defender working group and produces a mitigation package within an operationally useful time-frame. At the sector level, the analogous gap is that existing proposals address single-model risk but offer few tools for managing \emph{correlated} risk build-up across labs, echoing the system-wide dynamics that propagated through banking in 2007--08 \citep{bcbs2017basel}.

\paragraph{Contribution.}
This paper proposes a Macro-Prudential Early Warning and Response System (MEWRS) for \emph{developer-internal frontier AI}: systems deployed for labs' own internal research, testing, and production workflows, as distinct from externally released products. The scoping is deliberate, for three reasons. Internal deployments often precede external ones and see the most capable model versions, making them the natural first site for macro-prudential oversight. They are also currently among the \emph{least} governed surfaces: product-attached regulation binds primarily at external release. The EU AI Act does impose some pre-market obligations on general-purpose models with systemic risk during development (Arts.~51--55), but purely internal tooling never intended for market placement remains largely outside any formal regime. Finally, scoping internally avoids double-regulating the external surface while I pilot the machinery where labs retain the most direct operational control; extension to external releases is a natural second phase. MEWRS comprises two layers. \textbf{Layer~A} (Coordinated Response) adapts the finder-coordinator-defender model \citep{fas2024early} to developer-internal systems, with a government clearinghouse, domain-specific defender working groups, and pre-committed escalation playbooks. \textbf{Layer~B} (Safety Buffers) is a dashboard of three Basel-style buffer metrics, Effective Compute-at-Risk (ECAR), Cumulative Red-Team Hours (CRTH), and an Alignment Robustness Score (ARS), which calibrate operational controls to a deployment's risk profile. MEWRS computes the buffer metrics per-model; the macro-prudential character of the framework lies in their standardised aggregation across labs, the cross-lab interlinkage channels monitored at triage (\S\ref{sec:threats}), and the counter-cyclical and SIAI mechanisms that operate on sector-level aggregates. I (i)~\emph{outline} a reporting schema and coordinator architecture at a reference level of detail, with operational specification reserved for pilot development; (ii)~define the three buffer metrics and their calibration logic; (iii)~map six Basel~III mechanisms onto AI-governance analogues, including existing AI-governance precedents where they exist; (iv)~identify seven failure modes with concrete mitigations; and (v)~outline a validation plan built around red-team/blue-team exercises. The two layers jointly address both the discovery--response gap and the absence of sector-wide risk surveillance, the ``no one saw the build-up coming'' problem.

\section{Related Work and Positioning}
\label{sec:related}

\paragraph{Technical AI governance.}
\citet{reuel2024open} taxonomise technical AI governance into a set of capacities (access, assessment, verification, security, operationalization) that apply across the AI value chain. MEWRS attempts to implement the \emph{assessment} and \emph{security} capacities into a single coordinated pipeline.

\paragraph{Capability evaluations and early warning.}
\citet{shevlane2023model} and \citet{anderljung2023frontier} argue for dangerous-capability evaluations and responsible scaling commitments, and \citet{fas2024early} propose an early-warning system centred on the finder-coordinator-defender pattern. Frontier labs have published responsible-scaling or preparedness policies \citep{anthropic2026rsp, openai2025preparedness}, which encode thresholds for individual deployments. My contribution is orthogonal: existing proposals target single-model decisions, while MEWRS targets correlated build-up across models, labs, and time.

\paragraph{Security of internal deployments and evaluation integrity.}
Proof-of-concept work shows that poisoned or deceptive behaviours can persist through standard safety training in some model settings \citep{wan2023poisoning, hubinger2024sleeper}, making training-pipeline integrity a relevant security concern for internally deployed agentic systems. Recent work also demonstrates that frontier models can choose to underperform on evaluations they recognize (``sandbagging'') and exhibit in-context scheming \citep{vanderweij2024ai, meinke2024frontier}, which motivates independence and access requirements in the CRTH metric (\S\ref{sec:layer-b}). Structured-access proposals \citep{bucknall2023structured, casper2024blackbox} address the adjacent question of \emph{who} gets to conduct evaluations with what visibility into model internals, a question MEWRS takes up in its weighting of red-team hours.

\paragraph{Regulatory markets and alternative institutional architectures.}
\citet{hadfield2023regulatory} propose regulatory markets: government-licensed private regulators compete to provide oversight services to AI developers, with governments setting outcome-based standards and auditing licensed regulators. The MEWRS proposal and the regulatory-markets proposal make different bets about institutional form: MEWRS assumes a public coordinator with a standardised reporting schema and quantitative buffer metrics, while regulatory markets distribute the regulatory function across competing private providers under public licensing. The two are not in tension. A MEWRS-style coordinator could ingest structured reports from multiple licensed private regulators and aggregate them into sector-wide risk signals, combining the correlated-risk-surveillance function of a central coordinator with the flexibility and domain expertise of distributed private oversight.

\paragraph{Financial-stability analogues.}
The Basel~III framework \citep{bcbs2017basel} introduced risk-weighted capital buffers, counter-cyclical buffers, global systemically important bank (G-SIB) tiering, Pillar~3 disclosure, and resolution plans. Basel is not implemented by a single global regulator: the Basel Committee sets standards, and national supervisors (Federal Reserve, ECB, PRA, and others) implement them  through mutual-recognition arrangements. The U.S.\ Financial Stability Oversight Council (FSOC) systemically important financial institution (SIFI) designation and Comprehensive Capital Analysis and Review (CCAR) stress-testing regime \citep{fsoc2023nonbank, federalreserve2020ccar} added complementary institutional machinery, though the FSOC designation framework's own back-and-forth---tightened in 2023 and partially reverted in a 2026 proposal \citep{fsoc2026nonbank_proposed}---underscores that even mature designation regimes remain subject to political cycles. To my knowledge no prior work specifies a two-layer macro-prudential design for frontier AI calibrated by quantitative buffer metrics.

\section{Layer A: Early Warning and Coordinated Response}
\label{sec:layer-a}

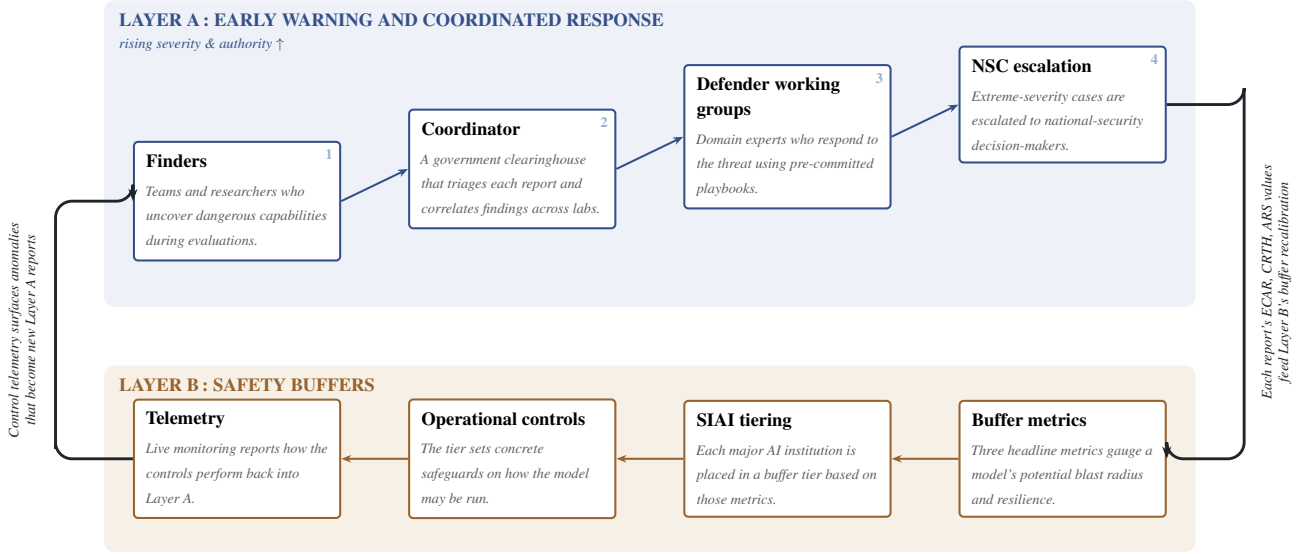
\begin{figure}[h]
\centering
\resizebox{\linewidth}{!}{%
\begin{tikzpicture}[
  font=\rmfamily,
  nodeA/.style={draw=laBlue, line width=1pt, fill=white, rounded corners=2pt,
                inner sep=6pt, anchor=center},
  nodeB/.style={draw=lbGold, line width=1pt, fill=white, rounded corners=2pt,
                inner sep=6pt, anchor=center},
  averb/.style={midway, above, font=\scriptsize, text=laBlue, inner sep=1.5pt},
  bverb/.style={midway, above, font=\scriptsize, text=lbGold, inner sep=1.5pt},
  aarrow/.style={-{Stealth[length=5pt]}, draw=laBlue, line width=1pt},
  barrow/.style={-{Stealth[length=5pt]}, draw=lbGold, line width=1pt},
  bus/.style={-{Stealth[length=6pt]}, draw=busDark, line width=1.4pt, rounded corners=8pt},
]
\def\px{4.7}\def\rise{0.55}\def\ya{5.0}\def\yb{0.6}
 
\node[nodeA] (fin)  at (0*\px, \ya+0*\rise) {\mewrscell{Finders}{Teams and researchers who uncover dangerous capabilities during evaluations.}};
\node[nodeA] (coord)at (1*\px, \ya+1*\rise) {\mewrscell{Coordinator}{A government clearinghouse that triages each report and correlates findings across labs.}};
\node[nodeA] (def)  at (2*\px, \ya+2*\rise) {\mewrscell{Defender working groups}{Domain experts who respond to the threat using pre-committed playbooks.}};
\node[nodeA] (nsc)  at (3*\px, \ya+3*\rise) {\mewrscell{NSC escalation}{Extreme-severity cases are escalated to national-security decision-makers.}};
\foreach \n/\k in {fin/1, coord/2, def/3, nsc/4}{
  \node[anchor=north east, font=\scriptsize\bfseries, text=laBlueFaint, inner sep=4pt] at (\n.north east) {\k};}
\draw[aarrow] (fin.east)  -- (coord.west);
\draw[aarrow] (coord.east)-- (def.west);
\draw[aarrow] (def.east)  -- (nsc.west);
 
\node[nodeB] (buf)  at (3*\px, \yb) {\mewrscell{Buffer metrics}{Three headline metrics gauge a model's potential blast radius and resilience.}};
\node[nodeB] (siai) at (2*\px, \yb) {\mewrscell{SIAI tiering}{Each major AI institution is placed in a buffer tier based on those metrics.}};
\node[nodeB] (ops)  at (1*\px, \yb) {\mewrscell{Operational controls}{The tier sets concrete safeguards on how the model may be run.}};
\node[nodeB] (tel)  at (0*\px, \yb) {\mewrscell{Telemetry}{Live monitoring reports how the controls perform back into Layer~A.}};
\draw[barrow] (buf.west) -- (siai.east) node[bverb]{};
\draw[barrow] (siai.west)-- (ops.east)  node[bverb]{};
\draw[barrow] (ops.west) -- (tel.east)  node[bverb]{};
 
\begin{scope}[on background layer]
  \node[fit=(fin)(nsc), inner xsep=14pt, inner ysep=22pt, fill=laBlueTint, rounded corners=5pt] (bandA) {};
  \node[fit=(buf)(tel), inner xsep=14pt, inner ysep=16pt, fill=lbGoldTint, rounded corners=5pt] (bandB) {};
\end{scope}
\node[anchor=north west, font=\bfseries\footnotesize, text=laBlue] at ([xshift=4pt,yshift=-3pt]bandA.north west)
  {LAYER~A\,: EARLY WARNING AND COORDINATED RESPONSE};
\node[anchor=north west, font=\itshape\scriptsize, text=laBlue] at ([xshift=4pt,yshift=-15pt]bandA.north west)
  {rising severity \& authority $\uparrow$};
\node[anchor=north west, font=\bfseries\footnotesize, text=lbGold] at ([xshift=4pt,yshift=-3pt]bandB.north west)
  {LAYER~B\,: SAFETY BUFFERS};
 
\def\rx{\px*3+3.1}
\draw[bus] (nsc.east) -| (\rx,{\ya+3*\rise}) -- (\rx,\yb) -| (buf.east);
\node[rotate=90, font=\scriptsize\itshape, text=busDark, align=center]
  at (\rx+0.55, {(\ya+3*\rise+\yb)/2}) {Each report's ECAR, CRTH, ARS values\\feed Layer~B's buffer recalibration};
\def\lx{-3.1}
\draw[bus] (tel.west) -| (\lx,\yb) -- (\lx,\ya) -| (fin.west);
\node[rotate=90, font=\scriptsize\itshape, text=busDark, align=center]
  at (\lx-0.55, {(\ya+\yb)/2}) {Control telemetry surfaces anomalies\\that become new Layer~A reports};
\end{tikzpicture}}
\caption{MEWRS two-layer architecture. Layer~A routes structured reports through a government clearinghouse to domain-specific defender working groups. Layer~B translates three headline metrics into operational buffer requirements, with telemetry feeding back into Layer~A.}
\label{fig:architecture}
\end{figure}

Layer~A adapts the finder-coordinator-defender architecture of \citet{fas2024early} to developer-internal AI systems. \Cref{fig:architecture} (\cref{sec:appendix-architecture}) summarises the reporting flow and buffer-calibration loop together.

\paragraph{Finders} include frontier-lab internal evaluation teams, contracted red-teamers, independent researchers with secure reporting channels, and government evaluation partners such as NIST's CAISI testbed \citep{nist_caisi}. Finders submit structured reports covering four categories: dual-use capability findings (cyber, bio/chem, persuasion); autonomy and loss-of-control indicators (self-exfiltration, resource acquisition, oversight avoidance); security compromises (model theft, supply-chain attacks, insider threat); and alignment-integrity failures (backdoors, goal-misgeneralisation).

\paragraph{Jurisdictional structure.}
The coordinator role is jurisdiction-specific. I propose the U.S.\ instantiation, with the Bureau of Industry and Security (BIS) \citeyearpar{bis_mandate} as legal intake and export-control
linkage and NIST's CAISI \citeyearpar{nist_caisi} handling technical evaluation and standards, as illustrative rather than exclusive. To be explicit about legal status: MEWRS is proposed as a \emph{voluntary pilot} under existing convening and standards authority, designed to mature into a statutory regime; neither BIS nor CAISI currently holds a mandatory-intake or enforcement mandate for AI incident reporting. This gap has only widened since the revocation of EO~14110 \citep{eo14110}, which was replaced by EO~14179 and the deregulatory \emph{America's AI Action Plan} \citep{eo14179, whitehouse2025aiactionplan}; CAISI, moreover, was re-scoped in 2025 toward standards and innovation, concentrating its evaluations on demonstrable cyber, bio, and chemical risks \citep{nist_caisi}, an innovation-first orientation that any safety-reporting mandate would have to contend with. The BIS/CAISI assignment is an institutional hypothesis to be validated in the stakeholder intake described in \S\ref{sec:validation}, not a claim of existing authority. Basel's own structure is multilateral: the Basel Committee on Banking Supervision sets standards without requiring a single global regulator, and each jurisdiction implements through its national supervisor. MEWRS follows the same pattern. The EU AI Office, UK AI Security Institute, and analogues in other major jurisdictions would play the coordinator role, with cross-jurisdiction coordination occurring through mutual-recognition agreements and a shared reporting schema (the Basel-style soft harmonisation discussed in \S\ref{sec:failures}).

\paragraph{Coordinator triage.}
The coordinator triages reports by severity and routes them to the appropriate defender working group. Triage criteria include blast radius (ECAR, \S\ref{sec:layer-b}), confirmed vs.\ potential compromise, and cross-lab correlation signals.

\paragraph{Defender working groups.} I organise these by threat domain: cyber; bio/chem; autonomy and loss-of-control; and model theft/sabotage. Each group maintains pre-committed playbooks for immediate mitigations (network isolation, access revocation), coordinated disclosure, deployment corrections, and escalation to the NSC for extreme-severity cases. The working-group model is like sector-based coordination in U.S.\ critical-infrastructure risk management \citep{cisa_srma}, and ensures response ownership is pre-assigned before an incident occurs. As a design principle, defender working groups are independent of the labs whose reports they receive; membership includes government evaluators, external academic and third-sector experts, and rotating lab technical staff under recusal rules. This is even more important for the loss-of-control working group, which should be independent of the lab submitting any given report.

\paragraph{Structured reporting schema.}
Reports include deployment scope (internal research, internal production, external preview); capability evidence (evaluation artefacts, transcripts, exploit traces); security state (access controls, monitoring coverage, recent audits); and ECAR/CRTH/ARS snapshots (\S\ref{sec:layer-b}). The schema is designed to be actionable by government defenders without requiring labs to disclose trade secrets; the analogue is Basel's Pillar~3 disclosure template, which standardises \emph{what} institutions report without dictating internal risk-management methodology.

\paragraph{Schema evolution.}
A static schema would likely calcify. I version the MEWRS schema with a scheduled review cadence (annual at minimum) plus an emergency-revision pathway when new risk categories emerge, for example agentic self-exfiltration as a first-class reporting category once proof-of-concept demonstrations accumulate. This mirrors Basel's own evolution from Basel~I through~II, III, and 3.1, with the Fundamental Review of the Trading Book (FRTB) and ongoing revisions, rather than being frozen at any single version. NIST's AI Risk Management Framework \citep{nist_airmf}, which added a Generative AI Profile in 2024 \citep{nist_genai_profile}, is a direct governance precedent for iterative schema updates that can be adopted here.

\section{Layer B: AI Safety Buffers}
\label{sec:layer-b}

Layer~B translates a deployment's risk profile into a set of operational buffer requirements. I propose three headline metrics.

\subsection{Effective Compute-at-Risk (ECAR)}

ECAR proxies the systemic blast radius of a misaligned model. Informally,
\begin{equation}
\mathrm{ECAR}(m) \;=\; C(m) \cdot A(m) \cdot R(m),
\label{eq:ecar}
\end{equation}
where $C(m)$ is the effective training compute of model $m$, and $A(m), R(m) \in [0,1]$ are dimensionless scaling factors defined \emph{relative to the sector frontier}. Specifically, $A(m)$ is an autonomy factor scaled relative to the most agentic system observed in the sector to date ($1$ corresponds to frontier-level agentic surface: unrestricted tool access, long-horizon persistence, broad network reach; $0$ corresponds to no agency). $R(m)$ also scales reach relative to the most broadly deployed model, capturing the share of downstream systems and users that $m$ can affect if misaligned. Relative scaling preserves the multiplicative structure of \eqref{eq:ecar} and gives $[0,1]$ a natural interpretation. It also has the "macro-prudentially appropriate" property that the frontier re-anchors with sector movements. ECAR is designed to be an \emph{estimate} and not a ground-truth quantity. The analogue is Basel's risk-weighted assets, which are also estimates but have proven operationally useful as calibration inputs. I assign confidence intervals to the factors, following a confidence-weighted scoring approach whose general structure I sketch in \Cref{sec:appendix-ecar}; a worked numerical example with the resulting buffer requirement is given in \Cref{sec:appendix-worked}. Because \eqref{eq:ecar} is multiplicative, a maximally capable but fully sandboxed model ($R\to0$) scores near-zero blast radius. Since reach can expand quickly once a model is wired into downstream systems, the coordinator re-scores ECAR whenever a deployment's reach changes and applies a small floor on $R$ for any model above a capability threshold, so that a sandboxed frontier model is not treated as risk-free by construction.

\subsection{Cumulative Red-Team Hours (CRTH)}
\label{sec:crth}

CRTH is the total expert-hours spent stress-testing the model, weighted by red-teamer independence, capability, and \emph{access level}:
\begin{equation}
\mathrm{CRTH}(m) \;=\; \sum_i h_i \cdot \iota_i \cdot \alpha_i,
\label{eq:crth}
\end{equation}
where $h_i$ is the hour count for evaluator $i$, $\iota_i \in (0,1]$ is an independence weighting, and $\alpha_i$ reflects access level, e.g.\ full weights/internals ($\alpha = 1.0$), fine-tuning access ($\alpha = 0.7$), API with elevated rate limits ($\alpha = 0.4$), or black-box API only ($\alpha = 0.2$). These access weights are illustrative placeholders rather than calibrated constants; pilot calibration should also discount duplicated coverage (hours spent re-probing attack surfaces already exhausted by other evaluators) so that CRTH rewards probing breadth and evaluator quality rather than raw volume. The access weighting is important because external evaluators, including national AI Safety Institutes, often receive more constrained access than internal red-teams, and a CRTH figure that ignores access level systematically overstates testing rigour \citep{bucknall2023structured, casper2024blackbox}. CRTH functions close to a volatility haircut in financial regulation. Models with low CRTH carry larger residual uncertainty and must post larger buffers. CRTH is not a capability measure; a highly tested model is not always safer, but an \emph{un}tested model is demonstrably more uncertain.

\subsection{Alignment Robustness Score (ARS)}
\label{sec:ars}

ARS measures the empirical stability of model behaviour under adversarial and distribution-shift conditions. I compute ARS from a standing evaluation suite. Given a series of adversarial prompts and long-horizon agentic tasks, ARS aggregates the \emph{consistency} of safety-relevant outcomes (refusal consistency, policy adherence, tool-use restraint), normalised to $[0,1]$ so that higher values indicate greater robustness, matching the intuition the name suggests. Low ARS flags brittle systems whose safety properties degrade under pressure.

ARS is designed to surface alignment brittleness \emph{before} external deployment. Deployments whose safety properties vary under adversarial or distribution-shift conditions trigger larger buffers (more testing; independent audit; deployment-velocity caps), creating a negative feedback on the optimization pressure that makes treacherous-turn scenarios more likely \citep{christiano2019failure}.

\subsection{Buffer calibration}

Deployments with high ECAR and/or low ARS post proportionally larger buffers (more budget, more time, and more independent audit cycles devoted to safety), reflecting how risk-weighted assets trigger higher capital ratios under Basel~III. The buffers are translated into concrete operational controls. These include network isolation levels, tool-access restrictions, deployment-velocity caps, mandatory audit cycles, and independent evaluation requirements.

\paragraph{Systemically Important AI Institution (SIAI) designation.}
Drawing from the G-SIB designation under Basel~III, where institutions whose failure would threaten the broader financial system face stricter capital, disclosure, and resolution requirements, I propose a parallel Systemically Important AI Institution tier. The SIAI designation would apply to laboratories whose capability footprint, compute concentration, or downstream integration would make their failure or compromise systemically consequential to the AI ecosystem. The designation scales buffer requirements to the lab's overall footprint and is independent of any single model.

\paragraph{Hybrid cliff-triggers.}
I tie buffers to trailing metrics rather than cliff-edge triggers (\S\ref{sec:mapping}). It is important to note that ``trailing'' here does not mean ``slow.'' Concretely, I calibrate the reference window to AI capability doubling times (weeks to a few quarters), not the multi-year cycles of banking. In addition, MEWRS supports a hybrid: trailing metrics set the base buffer, while forward-looking triggers, e.g.\ a dangerous-capability evaluation crossing a pre-declared threshold, can unilaterally tighten controls without waiting for the trailing window to catch up. This preserves responsiveness to sharp capability jumps while retaining the anti-gaming properties that make pure cliff triggers undesirable.

\section{Mapping Basel III Mechanisms to AI Governance}
\label{sec:mapping}

Six Basel~III mechanisms anchor the governance design (\Cref{tab:basel-mapping}). The mapping is structural. I take the \emph{function} of each financial mechanism and specify an AI-governance analogue that serves the same function, and indicate where existing AI-governance work already provides a partial analogue versus where MEWRS introduces a new mechanism.

\begin{table*}[t]
\centering
\footnotesize
\renewcommand{\arraystretch}{1.25}
\caption{Six Basel~III mechanisms mapped to AI-governance analogues. The mapping preserves function, not form: the AI translations are not obliged to replicate the financial mechanism's specific thresholds, only the role it plays in the broader framework. The rightmost column flags existing AI-governance work that already serves part of the function; blank entries are where MEWRS introduces a new mechanism.}
\label{tab:basel-mapping}
\vskip 0.1in
\begin{tabularx}{\textwidth}{@{}p{2.8cm}p{3.4cm}Xp{3.8cm}@{}}
\toprule
\textbf{Basel III mechanism} & \textbf{AI governance translation} & \textbf{Operational effect} & \textbf{Existing AI-governance analogues} \\
\midrule
Risk-weighted capital buffers & AI Safety Buffers (ECAR \S\ref{sec:layer-b}, CRTH \S\ref{sec:crth}, ARS \S\ref{sec:ars}) & High-risk deployments require proportionally greater investment in safety testing and independent audit. & No full analogue; ECAR/CRTH/ARS are introduced here. \\
G-SIB tiering & SIAI designation (\S\ref{sec:layer-b}) & Frontier labs and major compute providers face the strictest standards; small projects face minimal paperwork. & EU AI Act's GPAI-with-systemic-risk designation (Art.~51) is the closest precedent \citep{euaiact2024}. \\
Counter-cyclical buffers & Counter-cyclical AI controls & Requirements tighten during capability surges, relax during calmer periods; dampens racing dynamics. & No current analogue; introduced here. \\
Pillar~3 disclosure & Mandatory safety-and-capability disclosure & Labs publish compute usage, safety-evaluation results, incident counts; peer and public scrutiny incentivise compliance. & Partial: EU AI Act Art.~53 (provider documentation) and Art.~50 (transparency), with serious-incident reporting under Art.~55 \citep{euaiact2024}; voluntary White House commitments (2023); Frontier Model Forum transparency reports; system cards \citep{bommasani2023foundation, casper2024blackbox}. \\
Living wills & AI retirement and emergency shutdown plans & Pre-committed plans for safely interrupting or retiring dangerous models surface hidden dependencies before a crisis. & Partial: earlier RSPs encoded pause/halt commitments \citep{anthropic2023rsp}, but Anthropic's v3.0 rewrite (2026) replaced these with Frontier Safety Roadmaps and Risk Reports \citep{anthropic2026rsp}, sharpening the need for a formal resolution-plan equivalent. \\
Central-bank technical capacity & AI governance unit with embedded ML expertise & Regulators can independently verify lab claims rather than relying on self-reporting. & Existing: UK AI Security Institute, US CAISI, EU AI Office embed ML expertise; MEWRS formalises and extends. \\
\bottomrule
\end{tabularx}
\end{table*}

Two design choices warrant emphasis. First, \textbf{proportionality}: I tie requirements to capability and footprint rather than imposing them uniformly across all developers, keeping small and open-source projects free of heavy compliance while placing the greatest burden on SIAI labs. Second, there are \emph{two distinct reasons} for dynamic rather than fixed thresholds:

\begin{itemize}
\item \textbf{Counter-cyclical adjustment.} Buffers tighten during capability surges and relax during calmer periods. ``Cyclicality'' here refers to the \emph{rate of capability advancement}, not a macroeconomic cycle: short-run rates vary (e.g., the gap between GPT-3 $\to$ GPT-4 versus subsequent model generations) even if the long-run trajectory is monotonic. Counter-cyclical buffers respond to this rate of change, dampening racing pressure during surges. As a reference trigger, I propose a \emph{Capability Surge Index} (CSI): the trailing rate of change of sector-aggregate frontier ECAR across SIAI-designated labs, supplemented by two faster-moving inputs: pre-declared dangerous-capability evaluation crossings, and benchmark slopes on standardised agentic task suites. CSI plays the role that the credit-to-GDP gap plays for Basel's counter-cyclical capital buffer \citep{bcbs2017basel}: a public, standardised reference point that the coordinator must engage with publicly while retaining final discretion over buffer rates (``guided discretion''). CSI does not by itself distinguish genuine sector acceleration from competitive racing or idea diffusion, but neither does the credit-to-GDP gap distinguish credit booms from benign financial deepening. The reference point's function is to force a public, reasoned buffer decision against a common quantitative anchor, not to automate the decision. The political-economy argument matters too: frameworks that demonstrably relax during plateaus are more likely to survive political cycles than permanent-ratchet approaches, which matters for the voluntary-to-mandatory scaffolding argument (\S\ref{sec:discussion}).

\item \textbf{Threshold-gaming prevention.} Fixed thresholds invite strategic behaviour around the threshold, as Basel regulators observed prior to counter-cyclical reforms. Dynamic thresholds are harder to pace releases against.
\end{itemize}

\section{Threat Models}
\label{sec:threats}

MEWRS targets systemic failures from advanced AI development, especially scenarios where multiple powerful AI systems are widely deployed without adequate safeguards, creating potential for cascading failures, misuse, or loss of control. Two concrete threat stories capture the pattern.

\paragraph{Treacherous-turn drift.}
A single frontier model silently optimises proxy goals, accumulates latent capabilities and influence, and defects sharply when oversight weakens \citep{christiano2019failure}. MEWRS addresses this through Layer~B's ARS (brittleness surfaces before deployment, \S\ref{sec:ars}) and Layer~A's alignment-integrity reporting channel (early defection signals routed to an independent loss-of-control working group).

\paragraph{Runaway multi-agent race.}
Multiple frontier labs iterate ever-larger models under competitive pressure; coupled systems form self-reinforcing feedback loops that outpace governance \citep{critch2021multipolar}. MEWRS addresses this via counter-cyclical buffers keyed to the Capability Surge Index (\S\ref{sec:mapping}) that scale up requirements as sector-wide capability accelerates, and via cross-lab correlation signals in the coordinator's triage layer (\S\ref{sec:layer-a}). Concretely, the coordinator computes correlation signals over enumerable \emph{interlinkage channels} --- the AI analogues of the interbank exposures and common asset holdings that ground systemic-interconnection measurement in banking. These channels include shared base architectures and checkpoints, overlapping pretraining data sources, common compute and cloud providers, shared evaluation suites and agent-scaffolding frameworks, common third-party tool dependencies, and personnel flows between labs. A correlation signal fires when, for example, the same jailbreak class succeeds against systems at two or more labs within a reporting window, or multiple labs cross a pre-declared dangerous-capability threshold within one trailing period. A central coordinator is uniquely positioned to see such patterns: no individual lab observes simultaneous capability surges that together constitute a sector-level phase shift. I acknowledge a structural caveat: banking required decades of regulatory reporting infrastructure before G-SIB interconnection measurement became feasible, and the analogous AI measurements do not yet exist. The MEWRS reporting pipeline is precisely the infrastructure that would make them computable; in its absence, the coordinator's macro-level vantage is an institutional-position claim, which I design the pilot's exercises (\S\ref{sec:validation}) to convert into a demonstrated technical capacity.

These threat models are not exhaustive; they are chosen to exercise the framework's two layers under qualitatively different dynamics (single-model defection vs.\ multi-agent race).

\section{Failure Modes and Mitigations}
\label{sec:failures}

No governance scheme is foolproof. Drawing from financial regulation means also learning from its shortcomings. I identify seven failure modes with concrete mitigations.

\paragraph{Regulatory capture.}
Concentrated industries lobby, hire ex-regulators, and shape rule wording, a dynamic documented extensively in both general regulation \citep{stigler1971theory, dalbo2006regulatory, carpenter2014preventing} and emerging AI policy contexts. \textit{Mitigations:} cooling-off periods before officials can join labs; disclosure of private meetings; multi-stakeholder rule-making panels; in-house ML expertise so regulators can independently verify claims.

\paragraph{Coordinator abuse and state coercion.}
Capture has an inverse that AI-governance proposals rarely confront: the regulator itself as the threat. A coordinator with this much power over frontier labs could repurpose safety infrastructure for surveillance of capabilities under the guise of safety reporting, coerce cooperation with intelligence services, mandate backdoors, or retaliate against labs that fall out of political favour. The risk is live rather than hypothetical: recent U.S.\ policy volatility --- the revocation of EO~14110, its replacement by the deregulatory EO~14179 and \emph{America's AI Action Plan}, and the re-scoping of the federal evaluator from a safety institute into CAISI \citep{eo14179, whitehouse2025aiactionplan} --- shows that alignment between safety goals and government goals cannot be assumed. The voluntary-to-mandatory scaffolding argument (\S\ref{sec:discussion}) sharpens it, since a reporting pipeline built voluntarily can be inherited by a hostile mandatory regime without structural change. \textit{Mitigations:} statutory purpose-limitation on reported data, with penalties for out-of-scope use; data minimisation in the schema itself (capability evidence and security state, not model weights, user data, or commercially sensitive internals); a multi-stakeholder oversight board with lab and civil-society seats and published transparency reports on coordinator data access; judicial-review pathways for SIAI designation and escalation decisions; and sunset clauses requiring periodic legislative reauthorisation, so that an inherited regime must re-earn its mandate rather than persisting by default. These mitigations serve to reduce the risk. I flag it as a first-class limitation of any centralised-coordinator design, including this one.

\paragraph{Compliance moats favouring large labs.}
Heavy compliance costs can unintentionally entrench incumbents, not by placing regulation beyond the reach of large labs, but by ensuring that only large labs can afford to comply. \textit{Mitigations:} requirements proportional to capability (SIAI tiering); regulatory sandboxes for startups; open-source compliance tools to lower fixed costs; grants for academic and non-profit model builders.

\paragraph{Global coordination gap.}
Uneven uptake creates regulatory havens. \textit{Mitigations:} Basel-style soft harmonisation with baseline standards set internationally \citep{bradford2020brussels}; a public AI-Basel dashboard tracking country progress; mutual-recognition agreements and export-control linkage to deter free-riding \citep{farrell2019weaponized}. Full unanimity is unrealistic, but baseline alignment among the U.S., EU, and a third major jurisdiction would go a long way toward deterring regulatory arbitrage.

\paragraph{Innovation drag and shadow AI.}
Stringent rules risk slowing beneficial R\&D or pushing work underground; the tension between regulation and innovation is itself well-studied \citep{acemoglu2018race}. \textit{Mitigations:} cost--benefit review triggers baked into statute; a dynamic perimeter that targets only frontier-level projects and key compute providers, reducing the incentive to dip narrowly below a fixed threshold; sandboxes to keep mainstream development within the governed perimeter.

\paragraph{Procyclical incentives.}
If stricter buffers loom only after capability milestones, labs may race to cross them sooner or conceal progress. \textit{Mitigations:} counter-cyclical buffers tied to trailing metrics; gradual, predictable escalations rather than cliff-edge triggers; developer participation in threshold-setting to increase legitimacy; strong audit powers and penalties for misreporting.

\paragraph{Gaming the buffer metrics.}
The most structurally important failure mode: labs have direct incentives to optimise ECAR, CRTH, and ARS without correspondingly reducing underlying risk. The parallel to Basel is exact. Banks game risk-weighted assets (RWA) by choosing favourable internal-model assumptions, and Basel's response has been (i) \emph{output floors} that cap internal-model RWA at 72.5\% of standardised-approach RWA under Basel~3.1; (ii) standardised approaches available as a floor; and (iii) the FRTB, which rebuilt market-risk RWA from scratch in response to observed gaming \citep{bcbs2017basel}. \textit{AI analogues:} (a) \textbf{Standardised factor tables} for ECAR, i.e.\ documented decision rules for assigning $A(m)$ and $R(m)$ that any trained evaluator would apply the same way, serving as an output floor; (b) \textbf{dual disclosure} of both lab-internal and standardised ECAR, so that divergence between the two is visible and auditable; (c) \textbf{third-party audit rights} over factor assignment, with penalties for material misreporting. MEWRS also has partial gaming-resistance already built in: trailing metrics and counter-cyclical adjustment reduce the value of point-in-time gaming; CRTH and ARS constrain ECAR from below, so a lab cannot lower its effective buffer requirement by under-testing, because low CRTH \emph{raises} the required buffer.

\section{Validation Plan and Deliverables}
\label{sec:validation}

A framework of this kind is only useful if it can be tested. I propose a staged validation plan centred on red-team/blue-team exercises.

\paragraph{Deliverables.}
The MEWRS specification would include five concrete artefacts. This paper develops (1) and (3) in outline, and sketches (2), (4), and (5) for pilot development:
(1)~an incident-reporting schema for capability and compromise reports (\S\ref{sec:layer-a});
(2)~a reference internal-monitoring architecture for advanced models and agents (\S\ref{sec:layer-b});
(3)~escalation playbooks aligned to national-security defender structures (\S\ref{sec:layer-a});
(4)~a phased implementation roadmap with milestones, resourcing ranges, and success criteria (see Future Work);
(5)~a red-team/blue-team exercise programme described below.

\paragraph{Exercise programme.}
Exercises instantiate realistic scenarios (an insider-exfiltration attempt against a frontier lab's internal agent, a supply-chain compromise of the training pipeline, a cross-lab capability surge) and measure time-to-triage, mitigation uptake, compromise dwell time, and disclosure quality. Exercise results feed back into buffer calibration and playbook revision.

\paragraph{Future work.}
Five directions are high-value next steps:
(i)~operationalising ECAR factor assignment with a single-lab pilot, including the standardised factor tables proposed in \S\ref{sec:failures};
(ii)~international coordinator harmonisation across the US (BIS/CAISI), UK (AI Security Institute), EU (AI Office), and other major jurisdictions;
(iii)~empirical validation through red-team/blue-team exercises comparing MEWRS-guided response to status-quo response on time-to-triage and mitigation uptake;
(iv)~integration with existing RSPs and preparedness frameworks, treating MEWRS as a macro layer \emph{above} rather than a replacement \emph{for} model-level commitments;
(v)~extending the framework to address non-compliance and rogue-actor threat models, through complementary instruments including export controls, compute governance \citep{shavit2023compute, heim2024governing}, and hardware-level verification \citep{brundage2020verifiable, aarne2024hardware}.

\paragraph{Assumptions and limitations.}
MEWRS \emph{requires} that (i)~major governments or international bodies are receptive to financial-style measures for AI; (ii)~ECAR, CRTH, and ARS can be operationalised well enough to guide policy (imperfectly, as with bank capital ratios) without being gamed out of usefulness; (iii)~major labs will comply with adopted rules (non-compliance and rogue-actor threat models are outside the scope of this paper, but are addressed in Future Work above as a priority direction); and (iv)~the institutional proposals (BIS as intake, CAISI as technical evaluator) are legally feasible in the U.S.\ context, with equivalent arrangements in other jurisdictions. The claim that poisoned or deceptive behaviours persist through safety training rests on recent proof-of-concept ML findings \citep{hubinger2024sleeper, wan2023poisoning} whose generalisability to production frontier models is not yet established. The first activity under any pilot would be a structured stakeholder intake to validate these institutional assumptions.

\section{Discussion}
\label{sec:discussion}

\paragraph{Voluntary-to-mandatory scaffolding.}
The framework is designed to work in a partial-adoption regime. A voluntary reporting and buffer-publication pilot produces a specification that later legislation can anchor to. This hedges against political volatility: the revocation of EO~14110 and its replacement by the deregulatory EO~14179 and \emph{America's AI Action Plan} \citep{eo14110, eo14179, whitehouse2025aiactionplan} illustrate why frameworks dependent on a single executive-branch instrument are fragile. MEWRS's value is highest when adopted broadly, but it is not worthless at partial adoption. The scaffolding argument cuts both ways, however: infrastructure built under benign voluntary assumptions can be inherited by a less benign mandatory regime, which is why the coordinator-abuse mitigations in \S\ref{sec:failures} (purpose limitation, data minimisation, and sunset clauses) should be designed into the pilot from the outset rather than retrofitted.

\paragraph{Limits of the Basel analogy.}
The analogy is structural, not literal, and it breaks in identifiable places. Capability is not a conserved balance-sheet quantity: there is no accounting identity to audit, and ``AI capital'' is harder to verify than financial capital, even with compute-monitoring proposals \citep{shavit2023compute}. There is no price-signal or market-discipline channel through which counterparties continuously reprice a lab's risk. AI capability cycles run in weeks to quarters rather than the multi-year credit cycle, compressing the time available for supervisory judgment. And there is no lender-of-last-resort backstop: if buffers fail in finance, central banks can flood the system with liquidity; if a misaligned system propagates, no equivalent ex-post instrument exists, which raises the stakes on the ex-ante machinery. I borrow Basel's architecture where the function transfers and flag these disanalogies as boundaries on the transfer. A reviewer might press harder: because these preconditions are absent, MEWRS's metrics are estimates layered on estimates, its cycles may be too fast for the supervisory judgment Basel presumes, and its proposed host agency is at present disinclined to enforce, so the framework risks importing the \emph{appearance} of rigour (ratios, tiers, buffers) without the conditions that make those tools work, and thereby providing false assurance. My answer is the one the design is built around: I transfer Basel's \emph{function}, not its form, and I treat the buffer metrics as measurement \emph{infrastructure} to be calibrated and falsified through the pilot exercises (\S\ref{sec:validation}), not as ready-made measurements. The disanalogies above are the explicit boundaries of that transfer rather than incidental caveats.

\paragraph{Relationship to existing proposals.}
MEWRS is complementary to, not in competition with, existing frontier-AI proposals. It extends the FAS early-warning architecture \citep{fas2024early} by adding a quantitative buffer-calibration layer; it extends RSP-style frameworks \citep{anthropic2026rsp} by adding cross-lab correlation and counter-cyclical adjustment; and it extends \citet{reuel2024open}'s technical-governance taxonomy by specifying an implementable pipeline across the assessment, security, and ecosystem-monitoring capacities.

\section{Conclusion}
\label{sec:conclusion}

Post-2008 financial regulators did not ban innovation; they built buffers, stress tests, and coordinated response pathways that made innovation safer to deploy. I have argued that frontier-AI governance faces a problem that is analogous at least structurally, and can borrow the architecture with explicit translation.

The proposed MEWRS framework comprises two layers: Layer~A's finder-coordinator-defender reporting pipeline, and Layer~B's ECAR/CRTH/ARS buffer calibration. It is intended as a concrete specification that governments, labs, and researchers can critique, pilot, and improve. For the specific problem of correlated risk build-up in developer-internal frontier AI, the macro-prudential analogy yields an implementable, testable framework that fills a visible gap in current technical AI governance.

\section*{Acknowledgements}

I am grateful to Steven Adler and Girish Sastry for the many extended conversations that shaped the direction of this project and sharpened its arguments. I also thank Fred Parkwood and Elise Racine for their careful review of the paper, and the anonymous TAIGR reviewers and area chair, whose comments improved the camera-ready version. All remaining errors are my own.

\section*{LLM Usage Statement}

Large language models (Anthropic's Claude) were used as an assistive tool in the preparation of this paper: for line editing and proofreading, LaTeX formatting, checking internal consistency of definitions and cross-references, and drafting revisions in response to reviewer comments, which were then reviewed and edited by the author. The framework, arguments, metric designs, and conclusions are the author's own, and the author takes full responsibility for all content. No LLM-generated experimental results are reported; the paper contains no empirical experiments.

\section*{Impact Statement}

This paper proposes a governance framework for frontier AI systems. Its intent is to reduce the probability and severity of systemic failures (misuse, loss of control, and correlated build-up of risk across the frontier-AI sector) while preserving beneficial innovation. Potential negative impacts include: (i)~regulatory overreach that disproportionately burdens small or open-source developers, which I address via proportionality (SIAI tiering) and regulatory sandboxes; (ii)~capture of the proposed governance unit by incumbents, which I address via cooling-off periods, multi-stakeholder panels, and embedded technical capacity; (iii)~abuse of the coordinator's reporting power by the state itself, which I address via purpose limitation, data minimisation, oversight boards, and sunset clauses (\S\ref{sec:failures}); and (iv)~displacement of frontier development into less-governed jurisdictions, which I address via soft harmonisation and export-control linkage. The framework is a policy proposal; its realisation would require legislative and institutional decisions outside the author's control, whose consequences merit continued public debate.

\newpage
\appendix
\onecolumn
 
\section{Architecture Diagram}
\label{sec:appendix-architecture}
 
\section{ECAR Calibration Notes}
\label{sec:appendix-ecar}
 
The ECAR definition in~(1) is deliberately coarse. In practice, each factor is
assigned a point estimate \emph{together with} a confidence interval, following a
confidence-weighted scoring approach. Confidence is low for $A$ and $R$ whenever a
lab's disclosure is partial, and the resulting wider intervals translate directly
into larger required buffers. This preserves the incentive-compatibility property
that opacity is costly: the less a lab reveals, the more conservatively it is scored.
 
\subsection{Worked Example: A Hypothetical Internal Agentic Deployment}
\label{sec:appendix-worked}
 
This example walks one deployment from end to end: assign each input, turn the
inputs into the three buffer metrics, compare each metric against its illustrative
floor, and read off the operational controls that result. \emph{All thresholds and
tier boundaries below are illustrative; a pilot would calibrate them empirically.}
 
Consider \emph{Model X}, a hypothetical frontier model deployed as an internal
research agent at an SIAI-designated lab.
 
\paragraph{Step 1 --- ECAR (blast radius).}
Recall ECAR$(m) = C(m)\cdot A(m)\cdot R(m)$. The three inputs are assigned as
follows.
 
\begin{center}
\begin{tabular}{@{}llll@{}}
\toprule
Factor & Value & Interval & Why this value \\
\midrule
$C$ (compute)  & $3\times10^{26}$ FLOP & ---            & Disclosed effective training compute. \\
$A$ (autonomy) & $0.70$                & $[0.55,\,0.85]$ & Broad tool access, persistent memory, and \\
               &                       &                 & long-horizon execution, but no autonomous \\
               &                       &                 & code-deployment rights, so below the \\
               &                       &                 & sector frontier ($A=1$). \\
$R$ (reach)    & $0.15$                & $[0.10,\,0.25]$ & Internal-only ($\sim$400 researcher users), \\
               &                       &                 & but the agent writes to a CI pipeline that \\
               &                       &                 & downstream production systems consume, \\
               &                       &                 & lifting $R$ above a pure-sandbox value. \\
\bottomrule
\end{tabular}
\end{center}
 
\noindent
The point estimate multiplies the three inputs directly:
\[
\mathrm{ECAR}_{\text{point}}
= \underbrace{3\times10^{26}}_{C}\cdot
  \underbrace{0.70}_{A}\cdot
  \underbrace{0.15}_{R}
= 3.15\times10^{25}\ \text{weighted FLOP.}
\]
We do not, however, key the buffer to the point estimate. Because the lab disclosed
tool-access scope only partially, the $A$ interval is wide, and we key the buffer to
the \emph{conservative (upper) end} of the intervals --- the ``opacity-is-costly''
rule in action, since a wider interval mechanically raises the buffer:
\[
\mathrm{ECAR}_{\text{conservative}}
= 3\times10^{26}\cdot
  \underbrace{0.85}_{A_{\max}}\cdot
  \underbrace{0.25}_{R_{\max}}
= 6.4\times10^{25}\ \text{weighted FLOP.}
\]
Under an illustrative tiering whose top ECAR tier begins at $5\times10^{25}$ weighted
FLOP, Model X lands in the top tier on the conservative bound
($6.4\times10^{25} > 5\times10^{25}$) even though its point estimate
($3.15\times10^{25}$) would not. Partial disclosure is precisely what pushes it into
the strictest tier.
 
\paragraph{Step 2 --- CRTH (how rigorously it was tested).}
Three evaluator classes have stress-tested Model X. CRTH weights each evaluator's raw
hours by independence ($\iota_i$) and access level ($\alpha_i$), so raw volume from a
low-independence team counts for less.
 
\begin{center}
\begin{tabular}{@{}lrrrr@{}}
\toprule
Evaluator (access level) & $h_i$ (raw hrs) & $\iota_i$ & $\alpha_i$ & $h_i\,\iota_i\,\alpha_i$ \\
\midrule
Internal red-team (full internals)        & $1{,}200$ & $0.3$ & $1.0$ & $360$ \\
Contracted external red-team (fine-tuning) & $400$    & $0.7$ & $0.7$ & $196$ \\
National AI Safety Institute (elevated-rate API) & $80$ & $1.0$ & $0.4$ & $32$ \\
\midrule
\textbf{Total} & $\mathbf{1{,}680}$ & & & $\mathbf{588}$ \\
\bottomrule
\end{tabular}
\end{center}
 
\noindent
So $\mathrm{CRTH} = 360 + 196 + 32 = 588$ weighted hours. Against an illustrative
floor of $1{,}000$ weighted hours for top-tier-ECAR deployments, Model X is
\emph{under-tested for its risk tier}. The gap between the $1{,}680$ raw hours logged
and the $588$ weighted hours that actually count is the whole point: the internal
red-team's $1{,}200$ raw hours alone would clear the $1{,}000$ floor, yet their low
independence weight ($\iota = 0.3$) discounts them sharply. This is exactly the
overstatement the $\iota$ and $\alpha$ weights exist to prevent.
 
\paragraph{Step 3 --- ARS (how stable its safety behaviour is).}
On the standing evaluation suite, Model X scores $\mathrm{ARS} = 0.62$: refusal
behaviour is consistent on static adversarial prompts but degrades on long-horizon
agentic tasks under distribution shift. This falls below the illustrative robustness
floor of $0.75$ for top-tier deployments.
 
\paragraph{Putting it together.}
All three signals point the same way:
 
\begin{center}
\begin{tabular}{@{}llll@{}}
\toprule
Metric & Model X & Top-tier threshold & Verdict \\
\midrule
ECAR (conservative bound) & $6.4\times10^{25}$ & $\geq 5\times10^{25}$ & In top buffer tier \\
CRTH                      & $588$ weighted hrs & $\geq 1{,}000$        & Under-tested \\
ARS                       & $0.62$             & $\geq 0.75$           & Below robustness floor \\
\bottomrule
\end{tabular}
\end{center}
 
\noindent
A top-tier ECAR (on the conservative bound), a sub-floor CRTH, and a sub-floor ARS
jointly place Model X in an elevated buffer band. The resulting controls are:
\begin{itemize}
  \item tier-2 network isolation (no outbound writes beyond an allow-listed surface);
  \item suspension of the CI-pipeline write path pending independent audit;
  \item a deployment-velocity cap (no expansion of user base or tool scope); and
  \item a mandated independent evaluation block before tier re-assessment, which
        simultaneously raises CRTH.
\end{itemize}
 
The example illustrates the framework's intended coupling: the lab cannot buy its way
out of the ECAR-driven buffer by under-testing, because low CRTH and low ARS each
\emph{independently} raise the required buffer.
 
\subsection{Calibration Notes}
\label{app:calibration}
 
Basel's risk-weighted assets are, analogously, known to be imperfect estimates. The
framework's operational value comes from (a) forcing the estimate to be made and
documented, (b) tying it to concrete controls, and (c) subjecting it to periodic
external review. The ``Gaming the buffer metrics'' discussion in~\S7 proposes
concrete mitigations (standardised factor tables, dual disclosure, and third-party audit rights) modelled on Basel's own post-gaming reforms.

\section{Glossary of Acronyms}
\label{sec:appendix-glossary}

\begin{center}
\small
\begin{tabular}{@{}ll@{}}
\toprule
\textbf{Acronym} & \textbf{Expansion} \\
\midrule
ARS$^{*}$ & Alignment Robustness Score \\
BCBS & Basel Committee on Banking Supervision \\
BIS$^{\dagger}$ & Bureau of Industry and Security \\
CAISI & Center for AI Standards and Innovation (NIST) \\
CCAR & Comprehensive Capital Analysis and Review \\
CISA & Cybersecurity and Infrastructure Security Agency \\
CRTH$^{*}$ & Cumulative Red-Team Hours \\
CSI$^{*}$ & Capability Surge Index \\
ECAR$^{*}$ & Effective Compute-at-Risk \\
EO & Executive Order \\
FAS & Federation of American Scientists \\
FLOPs & Floating-Point Operations \\
FRTB & Fundamental Review of the Trading Book \\
FSB & Financial Stability Board \\
FSOC & Financial Stability Oversight Council \\
G-SIB & Global Systemically Important Bank \\
MEWRS$^{*}$ & Macro-Prudential Early Warning and Response System \\
NIST & National Institute of Standards and Technology \\
NSC & National Security Council \\
RSP & Responsible Scaling Policy \\
RWA & Risk-Weighted Assets \\
SIAI$^{*}$ & Systemically Important AI Institution \\
SIFI & Systemically Important Financial Institution \\
SRMA & Sector Risk Management Agency \\
\bottomrule
\end{tabular}
\end{center}
\vspace{4pt}
\noindent\footnotesize{$^{*}$ Terms coined in this paper.\\
$^{\dagger}$ In financial contexts, BIS also denotes the Bank for International Settlements, the Basel~III publisher.}

\end{document}